\documentclass[reprint,
 amsmath,amssymb,
 aps,
floatfix,
showkeys
]{revtex4-1}

\usepackage[USenglish]{babel}
\usepackage{slashed}
\usepackage{physics}
\usepackage{graphicx}
\usepackage{dcolumn}
\usepackage{bm}
\usepackage[usenames]{color}
\usepackage{xcolor}
\usepackage{float}

\newcommand{\be}{\begin{equation}}
\newcommand{\ee}{\end{equation}}
\newcommand{\bea}{\begin{eqnarray}}
\newcommand{\eea}{\end{eqnarray}}
\newcommand{\beas}{\begin{eqnarray*}}
\newcommand{\eeas}{\end{eqnarray*}}

\begin{document}

\title{Renormalons in a scalar self interacting theory: thermal, thermomagnetic and thermoelectric corrections for all values of temperature}


\author{M. Loewe$^{1,2,3}$ and R. Zamora$^{4,5}$ }
\affiliation{%
$^1$Instituto de F\'isica, Pontificia Universidad Cat\'olica de Chile, Casilla 306, Santiago 22,Chile.\\
$^2$Centre for Theoretical and Mathematical Physics, and Department of Physics, University of Cape Town, Rondebosch 7700, South Africa.\\
$^3$Centro Cient\'ifico-Tecnol\'ogico de Valpara\'iso CCTVAL, Universidad T\'ecnica Federico Santa Mar\'ia, Casilla 110-V, Valpara\'iso, Chile.\\
$^4$Instituto de Ciencias B\'asicas, Universidad Diego Portales, Casilla 298-V, Santiago, Chile.\\
$^5$Centro de Investigaci\'on y Desarrollo en Ciencias Aeroespaciales (CIDCA), Fuerza A\'erea de Chile, Casilla 8020744, Santiago, Chile.}%


\begin{abstract}
In this article we revisit the discussion of renormalons in the frame of a scalar self-interacting $\lambda \phi^{4}$ theory in the presence of thermomagnetic effects, i.e magnetic and thermal effects. Our results for the evolution of the residues is now given by an explicit analytic expression, valid for all values of temperature, without the necessity of separating the discussion in a low- and a high temperature region analyses. We carry out the same discussion for the case of an external constant electric field, obtaining also in this case an analytic expression for the whole range o temperature. In both cases, the location of the poles in the Borel plane does not change with respect to the vacuum case. Their residues, however, acquire a dependence on temperature and the external field. Results are presented for the evolution of residues in the thermomagnetic and thermoelectric cases. We show a comparison   with our previous results in the thermomagnetic case, presenting  also in detail the mathematical techniques needed for our analytic expressions to be valid in the whole range of temperature.

\end{abstract}


\maketitle

\section{Introduction}\label{sec1}
We know since the work by Dyson on the validity of perturbative expansions in Quantum Electrodynamics (QED)  \cite{Dyson} that power series expansions in Quantum Field Theory (QFT), in general, are divergent objects. When going to high orders in perturbative expansions, the divergence grows like $n!$, where n is the order of expansion, and this is due, essentially, to the multiplicity of diagrams that contribute to Green functions at such expansion order. A way or procedure to improve the convergence relies on the Borel transform \cite{Altarelli,Rivasseau,Khanna}. However, in some cases even the Borel transformed series are divergent, spoiling  the meaning of the whole procedure. The singularities responsible for this divergent behavior are the renormalons. Beyond such a singularity, perturbation theory does not make any sense. For a  classical review, see \cite{Beneke}. Recently there has been a renewed interest in the subject by considering one loop renormalization group equation  in multi-field  theories \cite{Vasquez} or by considering finite temperature mass correction in the $\lambda \phi ^{4}$ theory, reanalyzing the temperature dependence of poles and their residues \cite{Cavalcanti}. Quite recently, the relation between renormalons and analytic trans-series has been considered in the context of field theories in two dimensions \cite{Marino}. It is also quite interesting the relation between resurgence and trans-series where it is possible to give sense to divergent perturbative series by invoking some additional information on the structure of coupling singularities in the complex plane \cite{Shifman}. The phenomenological relevance of trans-series in the resurgent framework has been stressed in a recent article on the muon $g-2$ anomaly through a discussion of infrared renormalons in the QCD Adler function \cite{Maiezza2}. The occurrence of renormalons in QFT has recently been discussed  from a quite general point of view, in fact without reference to Feynman diagrams.  A careful discussion of the renormalization group, recasting as a resurgent equation, shows clearly the presence of such singularities  \cite{v1}. The relation of renormalons to non-local effects has also been stressed in literature \cite{v2}. In a future article we plan to explore trans-series for perturbative expansions in QFT, in the presence of external effects like temperature and external fields.  Other sources of divergences, as instantons in quantum chromodynamics (QCD) \cite{Gross,Shafer}, will not be considered here.

\medskip
\noindent
In peripheral heavy ion collisions, huge magnetic fields appear \cite{warringa}. In fact, the biggest fields existing in nature. The interaction between the produced pions in those collisions may be strongly affected by the magnetic field.Temperature is, of  course, also present in such scenario. In fact, several studies on the behavior of different physical parameters in the presence of external magnetic field and/or temperature have been carried out by different authors. \cite{bali01,iranianos,zamora1,zamora2,zamora3,simonov03,aguirre02,tetsuya,dudal04,kevin,gubler,noronha01,morita,Ayala0,Ayala1,morita02,sarkar03,band,nosso1,nosso03,Ayala2,Ayala3,zamora4,pi1,pi2}. In the same way, if we consider now peripheral collisions  of asymmetric nuclei, for example Cu-Au collisions, we expect the appearance of a strong electric, dipole-type, field due to the imbalance between the number of positive charges in both nuclei. The relevance of the appearance of such an electric field has been considered in the discussion of the phase diagram of some QFT models as, for example,  a two-flavor Nambu-Jona-Lasinio model \cite{farias}. Quite recently, now in the context of a $\lambda\Phi^4$ theory, we have explored, through a discussion of the effective potential, the emergence of catalysis or inverse electric-catalysis for any value of temperature \cite{nosotros}.

\noindent In the present article we analyze, in the frame of a self interacting scalar  $\lambda \phi^{4}$ theory, the influence of a magnetic field and an electric field, taken separately, together with temperature on the position of the U.V. renormalons (the only relevant in $\lambda\phi^4$ theory) and their residues in the Borel plane.

Finally, we would like to mention  also other extensions as the q-Borel series allowing the discussion of series whose coefficients grow like $(k!)^q$ \cite{Cavalcanti2}. Also, there have been new attempts to find corrections to the Beta function in QED with $N_{f}$ flavors by considering closed chains of diagrams, like renormalons, and computing corrections of order $N _{f}^{-2}$ and $N_{f}^{-3}$ \cite{Dunne}. These authors found a new logarithmic branch cut whose physical role is not clear. Probably the same situation will occur in self interacting scalar theories with several components.  

\medskip
\noindent
This article is organized as follows. In section II  we present a brief introduction about  Borel summable series and poles in the Borel plane. In section III we present for completeness the calculation of the vacuum renormalon, i.e. in absence of temperature and/or external fields.  In section IV we present the pure thermal discussion of the renormalon, obtaining a single analytic expression valid for the whole range of temperature. Section V presents the case where temperature and magnetic field effects are included in the renormalon-type corrections to the propagator. Section VI presents the case that includes temperature and electric field effects. In section VII we show some numerical comparisons between the electric and the magnetic situation, both in the magnetic and in the electric field cases, being our expressions valid for the whole range of temperature. Finally in section VIII we present our conclusions.




\section{The Borel Transform} \label{Boreltransform}

We will briefly remind the Borel transform method  designed to make sense of potentially divergent series \cite{Altarelli}.
\\
For a divergent perturbative expansion 

\begin{equation}
D[a]=\sum_{n=1}^\infty D_na^n \text{,}
\end{equation}

\noindent
the Borel transform $B[b]$ of the series $D[a]$ is defined through 

\begin{equation}
B[b]=\sum_{n=0}^\infty D_{n+1}\frac{b^n}{n!} \text{.}
\end{equation}

The inverse transform is, 

\begin{equation}
D[a]=\int_0^\infty db \hspace{2mm} e^{-b/a} B[b] \text{.}
\end{equation}

We need the last integral to be convergent, being $B[b]$ free form singularities in the range of integration. If these conditions are fulfilled, we say that the original series $D[a]$ is Borel summable.

It is easy to check that all convergent series are also Borel summable. For the case of divergent series this is not necessarily the case. If we find poles in the $0-\infty $ range of integration of the previous equation, the series is no longer Borel summable. In this case, however, it is still possible to make sense to this integral through a prescription for the integration path in the complex $[b]$-plane, avoiding the pole. This will be, however, a prescription-dependent result.

A classical reference about divergent series is the book by Hardy, \cite{hardy}.

It is known that perturbative expansions in quantum field theory are not Borel summable. There are two sources for the appearance of singularities in the Borel plane: renormalons and instantons. Here we do not want to comment about the latter possibility. In QCD, renormalons have been a matter of debate since these objects may affect our understanding of the gluon condensate \cite{Beneke}.

\section{Renormalons in the vacuum}\label{secRenVacio}
 In the $\lambda \phi^{4}$ theory the renormalon type diagrams that produce poles in the Borel plane correspond to a correction to the two-point function

\vspace{0.5cm}
\begin{figure}[h] 
\begin{center} 
\includegraphics[width=7cm]{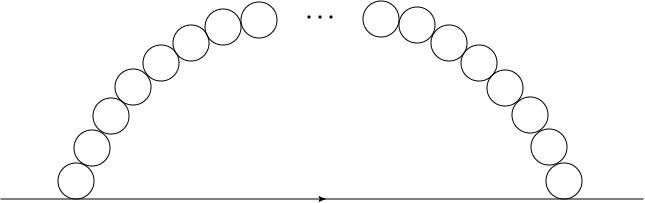}
\caption{Renormalon-type contribution to the two point function.}
\label{fig:diagrenormalon}
\end{center}
\end{figure}

First we remind the calculation of this diagram with the insertion of k bubbles, summing then over $k$ and studying the behavior of its transform in the Borel plane.

We will denote by $R_k(p)$ the diagram of order $k$ shown in Fig. {\ref{fig:diagrenormalon}}, where  $p$ is the four-momentum  entering and leaving the diagram and  $q$ is the four momentum that goes around the chain of bubbles

\begin{equation}\label{Rk}
R_k(p)=\int \frac{d^4q}{(2 \pi)^4}\frac{i}{(p+q)^2-m^2+i\epsilon}\frac{[B(q)]^{k-1}}{(-i\lambda)^{k-2}} \text{.}
\end{equation}

\begin{figure}[h]
\begin{center}
\includegraphics[scale=0.5]{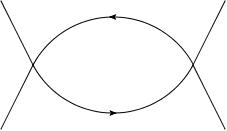} 
\caption{Fish-diagram.}
\label{fig:diagfish}
\end{center}
\end{figure}

In this expression,  $B(q)$ corresponds to the contribution of one bubble in the chain which is equivalent, of course, to the one-loop correction, in the s-channel, of the vertex, the so called   fish-diagram  (see Fig. \ref{fig:diagfish}). The factor $(-i\lambda)^{k-2}$ has been added to cancel the vertices that have been counted twice along the chain.

The expression for  $B(q)$ is a well known result \cite{bailin} 
\begin{equation}
\begin{split}
B(q)=\frac{(-i\lambda)^2}{2}\int  \frac{d^4k}{(2 \pi)^4}\frac{i}{k^2-m^2+i\epsilon} \\ \times\frac{i}{(k+q)^2-m^2+i\epsilon}
\end{split}
\end{equation}
\noindent given by
\begin{equation}\label{fishexact}
\begin{split}
B(q)=\frac{-i\lambda^2}{32\pi ^2} \left[ \Delta  - \int _0 ^1 dx \log \left( \frac{m^2 - q^2 x(1-x) - i\epsilon}{\mu ^2} \right) \right] 
\end{split}
\end{equation}

\noindent where $\Delta$ is the divergent part that will be canceled by counterterms in the renormalization procedure, being $\mu$ an arbitrary mass-scale  associated to the regularization procedure,  which always appear when we go through the renormalization program.

The contribution to the renormalon comes from the deep euclidean region in the integral in Eq.~(\ref{Rk}), i.e. where  $-q^2\gg m^2$. In this way,  $B(q)$ in Eq.~(\ref{fishexact}) can be approximated as:

\begin{equation}\label{fishDE}
B(q) \approx \frac{-i\lambda^2}{32\pi ^2} \log(-q^2/\mu^2) \text{,}
\end{equation}

Inserting this result in Eq.~(\ref{Rk}) and performing a Wick rotation, we find

\begin{eqnarray}
R_k(p)&=&\frac{-i\lambda^k}{(32\pi ^2)^{k-1}} \int \frac{d^4q}{(2 \pi)^4}\frac{1}{(p+q)^2+m^2} \nonumber \\
&\times& [\log(q^2/\mu^2)]^{k-1} \text{.}
\end{eqnarray}

This is an ultraviolet divergent expression. However, since the theory is renormalizable, we can separate this expression into  a finite and a divergent part. We are only interested in the finite part. For this we expand the propagator in powers of $1/q^2$, neglecting the first two terms that leave divergent integrals. We can do this because of the appearance of counterterms.

So, we find
\begin{align}\label{R2}
\begin{split}
R_k =& -i \left( \frac{\lambda}{32\pi ^2} \right)^k 4 \int _\Lambda ^\infty dq [\log (q^2/\mu^2)]^{k-1}q^3 \\
&\times \left[\frac{m^4}{q^6}-\frac{m^6}{q^8}+...\right]
\end{split}
\end{align}
with $\Lambda >0$.
The dependence on the external momentum $p$ has disappeared. The lower limit $\Lambda $ comes from the fact that we are interested in the deep euclidean region. It has to be fixed in order to fulfill this condition.

Introducing $q=\mu e^t$ in the first two terms of Eq.~(\ref{R2}), and fixing $\Lambda = \mu$, we find

\begin{align}
R_k =&  -i \left( \frac{\lambda}{32\pi ^2} \right)^k  \frac{4 m^4}{\mu^2}\int_0 ^\infty dt e^{-2t} (2t)^{k-1} \nonumber\\
&\times \left(1-\frac{m^2}{\mu^2} e^{-2t} \right)  \nonumber \\
=&  -i \left( \frac{\lambda}{32\pi ^2} \right)^k \frac{2 m^4}{\mu^2} \Gamma (k) +i \left( \frac{\lambda}{64\pi ^2} \right)^k \frac{2 m^6}{\mu^4} \Gamma (k) \nonumber\\
=& -i \left( \frac{\lambda}{32\pi ^2} \right)^k \widetilde{m}^2 \Gamma (k) +i \left( \frac{\lambda}{64\pi ^2} \right)^k \frac{m^2}{\mu^2}\widetilde{m}^2 \Gamma (k)
\end{align}
with $\widetilde{m}^2=2m^4/\mu^2$. We see that this diagram grows like $k!$, inducing then a pole in the Borel plane.\\

By taking the series $\Sigma R_k$,

\begin{equation}
D[\lambda] = \sum_k \frac{-i}{(32\pi ^2)^k}\Gamma(k)\lambda^k \text{,}
\end{equation}
its Borel transform is given by 
\begin{eqnarray}
B[b] &=& \sum_k \left( \frac{R_k}{\lambda^k} \right)\frac{b^{k-1}}{(k-1)!} \text{,}\nonumber \\
&=& -i \widetilde{m}^2\frac{1}{1-b/32\pi^2} +i \frac{m^2}{\mu^2}\widetilde{m}^2\frac{1}{1-b/64\pi^2} \text{,}
\end{eqnarray}

identifying, finally, the leading pole on the positive semi real axis in the Borel plane $b=32\pi^2$ and the second pole in $b=64\pi^2$.

\section{THERMAL
RENORMALONS}\label{thermorenormalons}
We are going to calculate the renormalon contribution to the two point function at finite temperature. In a previous article \cite{renormalontermomagnetico}  we went through this discussion but it was necessary to separate the high and low temperature cases. Now we have at our disposal new techniques that leave a closed analytic expression, valid for the whole temperature range, for the thermal corrections we are looking for. Temperature is introduced in the frame of the imaginary time formalism where bosonic Matsubara frequencies appear
\cite{LeBellac, Kapusta}
\begin{equation}
k_4 \rightarrow \omega_{n} = \frac{2\pi n}{\beta},\; n\in Z,
\end{equation} 
where $\beta=1/T$, and where the integral in $k_4$ converts into a sum according to,
\begin{eqnarray}
\int \frac{d^4k}{(2\pi)^4}f(k) \rightarrow \frac{1}{\beta} \sum_{n\in Z} \int \frac{d^3k}{(2\pi)^3} f(i\omega_n,k).
\end{eqnarray}
The diagram to be calculated is
\begin{equation}
R_{T,k}=\frac{1}{(-i\lambda)^{k-2}} \int \frac{d^4q}{(2\pi)^4} i D(p+q)[B^T(q)]^{k-1},
\end{equation}
Notice that in principle the $D(p+q)$ propagator could be temperature dependent. However, in our previous work \cite{malaquias} when we considered the pure magnetic corrections case, we showed that if the propagator $D(p+q)$ included magnetic effects, it´s contribution to  $B(q)$ is sub-leading.The same situation happens also here, when dealing only with temperature corrections. Therefore, our calculation will be carried out by considering only $B(q)$ as temperature dependent, whereas $D(p+q)$ will be handled without temperature effects. First let us consider one bubble 
\begin{equation}
B^T(q) =  \frac{(-i\lambda)^2}{2}  i T \sum_{n} \int \frac{d^3k}{(2\pi)^3}   D^T(k) D^T(k+q), \label{fishT}
\end{equation}
where 
\begin{equation}
  D^T(k) = \frac{1}{\omega_n^2+k^2+m^2},  
\end{equation}
therefore
\begin{eqnarray}\label{26}
B^T(q) &=&  \frac{(-i\lambda)^2}{2} i T\sum_{n} \int \frac{d^3k}{(2\pi)^3} \frac{1}{\omega_n^2+k^2+m^2} \nonumber \\
&\times&\frac{1}{(\omega_n+\omega)^2+(k+q)^2+m^2}. 
\end{eqnarray}
The sum over Matsubara frequencies is calculated using well knowing techniques \cite{LeBellac}, getting
\begin{align}
\begin{split}
B^{T}(q) =  \frac{(-i\lambda)^2}{2} i\int & \frac{d^3k}{(2\pi)^3} \frac{- s_1 s_2}{4 E_1 E_2} \\ &\times\frac{1+n(s_1 E_1)+n(s_2 E_2)}{i \omega -s_1 E_1 - s_2 E_2}, 
\end{split}
\end{align}
where 
\begin{center}
\begin{tabular}{lr}
$E_1^2=k^2+m^2$, 		& $s_1=\pm1$, \\
$E_2^2=(k+q)^2+m^2$, 	& $s_2=\pm1$, \\
\multicolumn{2}{c}{$n_i(E_i) = 1/(e^{E_i/T}-1)$}.\\
\end{tabular}
\end{center}

We obtain
\begin{align}
B^{T}(q) =&  \frac{(-i\lambda)^2}{2}i \int \frac{d^3k}{(2\pi)^3} \frac{1}{4 E_1 E_2} \nonumber\\
&\times\Biggl[ (1+n_1+n_2) \Biggl( \frac{1}{i\omega - E_1 - E_2} -\frac{1}{i\omega + E_1 + E_2} \Biggr) \nonumber\\  
&-(n_1-n_2)\Biggl( \frac{1}{i\omega - E_1 + E_2} -\frac{1}{i\omega + E_1 - E_2} \Biggr)\Biggr] \nonumber\\
\equiv& B_{\text{vac}}(q)+ B_T(q),
\end{align}
where
\begin{align}
\begin{split}
B_{\text{vac}}(q) =&  \frac{(-i\lambda)^2}{2}i \int \frac{d^3k}{(2\pi)^3} \frac{1}{4 E_1 E_2} \\
&\times \Biggl( \frac{1}{i\omega - E_1 - E_2} -\frac{1}{i\omega + E_1 + E_2} \Biggr), 
\end{split}
\end{align}

is the vacuum part equal to Eq.~(\ref{fishDE}) in the deep euclidean region. 
Notice that we will use the notation $B^T(q)$ for the total fish diagram, including both vacuum and the thermal corrections, whereas $B_T(q)$ will refer only to thermal corrections to the fish
\begin{eqnarray} \label{parteT}
&&B_T(q) =  \frac{(-i\lambda)^2}{2}i \int \frac{d^3k}{(2\pi)^3} \frac{1}{4 E_1 E_2} \nonumber \\
&\times&\Biggl[ (n_1+n_2) \Biggl( \frac{1}{i\omega - E_1 - E_2} -\frac{1}{i\omega + E_1 + E_2} \Biggr) \nonumber \\  
&-&(n_1-n_2)\Biggl( \frac{1}{i\omega - E_1 + E_2} -\frac{1}{i\omega + E_1 - E_2} \Biggr)\Biggr], \end{eqnarray}
being the temperature dependent part. Since the contribution to the renormalon comes from the deep euclidean region, we calculate Eq.~(\ref{parteT}) in the limit  $-q^2\gg m^2$, obtaining
\begin{eqnarray}
B_T(q) &\approx& \frac{i \lambda^2}{4 \pi^2 q^2} \int_0^\infty{dk} \frac{k^2}{\sqrt{k^2+m^2}}  \frac{1}{e^{\sqrt{k^2+m^2}/T}-1}. 
\end{eqnarray}
The previous integral can be obtained analytically \cite{nosotros}, getting
\begin{equation}
B_T(q) \approx \frac{i \lambda^2}{4 \pi^2 q^2} \sum_{n=1}^\infty \frac{m T}{n} K_1(nm/T)  
\end{equation}
where $K_1$ is the modified Bessel function of the second kind. Hence the contribution for one bubble is     
\begin{equation}
B^{T}(q) \approx \frac{-i\lambda^2}{32\pi ^2} \left( \log(q^2/\mu^2) -\frac{8}{q^2} \sum_{n=1}^\infty \frac{m T}{n} K_1(nm/T) \right).
\label{bubbleT}
\end{equation}
Now we have to insert this temperature dependent fish diagram term in the chain of bubbles that define the renormalon-type diagram. We find
\begin{eqnarray}
R_{T,k}&=&\frac{1}{(-i\lambda)^{k-2}} \int \frac{d^4q}{(2\pi)^4} i D(p+q)\left( \frac{-i \lambda^2}{32 \pi^2}\right)^{k-1} \nonumber \\
&\times& \left( \log(q^2/\mu^2) -\frac{8}{q^2}  \sum_{n=1}^\infty \frac{m T}{n} K_1(nm/T)   \right)^{k-1},
\end{eqnarray}
As we already mentioned, we proceed to expand the propagator $D(p+q)$ in powers of $1/q^2$ neglecting the first two terms that give rise to divergent integrals. We have then to calculate
\begin{eqnarray}
&&R_{T,k}=\frac{1}{(-i\lambda)^{k-2}} \int \frac{d^4q}{(2\pi)^4} \left[\frac{m^4}{q^6}-\frac{m^6}{q^8}+ \cdots \right] \nonumber \\
&\times& \left( \frac{-i \lambda^2}{32 \pi^2}\right)^{k-1} \Biggl( \log(q^2/\mu^2) \nonumber \\
&&-\frac{8}{q^2}  \sum_{n=1}^\infty \frac{m T}{n} K_1(nm/T)   \Biggr)^{k-1}.
\end{eqnarray}
Using the binomial theorem
\begin{equation}
    (A+B)^N=A^N+N \cdot A^{N-1} B + \cdots,
\end{equation}
we get
\begin{eqnarray}
&&R_{T,k}=-i \frac{\lambda^k}{(32 \pi^2)^{k-1}} \int  \frac{d^4q}{(2\pi)^4} \left[\frac{m^4}{q^6}-\frac{m^6}{q^8}+...\right] \nonumber \\
&&\times  \Biggl[\log(q^2/\mu^2)^{k-1} -(k-1)\frac{8}{q^2} \sum_{n=1}^\infty \frac{m T}{n} K_1(nm/T)  \nonumber \\ &&\times\log(q^2/\mu^2)^{k-2} + ... \Biggr].  
\end{eqnarray} 
Notice that the vacuum leading term as well as the next to leading order (NLO) vacuum term  can be extracted from the first two terms inside the first square parenthesis, together with the first term of the second square parenthesis in the previous equation. The leading term in the magnetic field strength is obtained by multiplying the first term of the first square parenthesis with the second term of the second square parenthesis in the above equation. The next terms are sub-leading.

In this way, following the same procedure of section (\ref{secRenVacio}) and performing the angular integrals we find

\begin{eqnarray}
&&R_{T,k} = -i 4m^4 \left( \frac{\lambda}{32\pi ^2}  \right) ^k \int dq \Biggl[\frac{\left(\log (q^2/\mu^2)\right)^{k-1}}{q^3}  \nonumber \\
&-&\frac{m^2}{q^5}\log (q^2/\mu^2)^{k-1} \nonumber \\
&-& \frac{(k-1)8  (\log(q^2/\mu^2))^{k-2}}{q^5} \sum_{n=1}^\infty \frac{m T}{n} K_1(nm/T)  + ... \Biggr]   \text{.} \nonumber \\
\end{eqnarray}

Using the change of variable  $q=\mu e^t$, $dq=\mu e^tdt$, 

\begin{eqnarray}
&&R_{T,k} = -i \left( \frac{\lambda}{32\pi ^2} \right)^k \frac{4m^4}{\mu^2} \int dt \Biggl[e^{-2t}(2t)^{k-1} \nonumber \\
&-& \frac{m^2}{\mu^2}e^{-4t} (2t)^{k-1} - \frac{(k-1)8 e^{-4t}(2t)^{k-2}}{ \mu^2} \nonumber \\
 &&\times \sum_{n=1}^\infty \frac{m T}{n} K_1(nm/T)  + ... \Biggr]  \text{,} \nonumber \\
\end{eqnarray}
we see the appearance of the Gamma function in both terms. Using the definition of $\widetilde{m}$, we have 

\begin{eqnarray}
&&R_{T,k} = -i \widetilde{m}^2 \left( \frac{\lambda}{32\pi ^2}  \right)^k \Gamma(k) +i \left( \frac{\lambda}{64\pi ^2} \right)^k \frac{m^2}{\mu^2}\widetilde{m}^2 \Gamma (k) \nonumber \\
 &-&2i \frac{\widetilde{m}^2}{\mu^2} \left( \frac{\lambda}{64\pi ^2}  \right) ^k 8 \sum_{n=1}^\infty \frac{m T}{n} K_1(nm/T)  \Gamma(k) + ... \hspace{1mm}  \text{,}\label{rT} 
\end{eqnarray}
where we have used also the property $\Gamma(z+1)=z\Gamma(z)$, Eq.~(\ref{rT}) can be written as
\begin{eqnarray}
R_{T,k}&\equiv& -i \widetilde{m}^2 \left( \frac{\lambda}{32\pi ^2}  \right)^k \Gamma(k) +i \left( \frac{\lambda}{64\pi ^2} \right)^k \frac{m^2}{\mu^2}\widetilde{m}^2 \Gamma (k) \nonumber \\
 &-&2i \frac{\widetilde{m}^2}{\mu^2} \left( \frac{\lambda}{64\pi ^2}  \right)^k \Gamma(k) F_{\text{T}} + ... \hspace{1mm}  \text{,}
\end{eqnarray}
where
\begin{equation}
F_{\text{T}}= -8\sum_{n=1}^\infty \frac{m T}{n} K_1\left(\frac{nm}{T}\right).  \end{equation}
Now we can find the Borel transform $B[b]$ of $\Sigma R_{T,k}$,
\begin{eqnarray}
B[b]&=&\sum_k \left( \frac{R_{T,k}}{\lambda ^k}  \right) \frac{b^{k-1}}{(k-1)!} \text{,}\nonumber \\
&=& - i\widetilde{m}^2 \sum_k \left( \frac{1}{32\pi ^2}  \right) ^k b^{k-1} +\left(\frac{i m^2}{\mu^2}\widetilde{m}^2-\frac{2i \widetilde{m}^2}{ \mu^2 } F_{\text{T}} \right) \nonumber \\
&\times&\sum_k \left( \frac{1}{64\pi ^2}  \right)^k b^{k-1} + ... . \nonumber \\
\end{eqnarray}
These sums correspond to well known geometrical series, obtaining then our final result
\begin{equation}
B[b] = \frac{-i \widetilde{m}^2}{b-32\pi ^2} +\Biggl[i \frac{m^2}{\mu^2}\widetilde{m}^2- \frac{2i \widetilde{m}^2}{ \mu^2}F_{\text{T}} \Biggr] \frac{1}{b-64\pi ^2} + ...  \text{.} 
\label{todoT}
\end{equation} 
The main difference with respect to what we found in \cite{renormalontermomagnetico} is that  Eq.~(\ref{todoT}) is valid for the whole range of temperature. The location of the poles in the Borel plane, as expected, does not depend on temperature. The residues, nevertheless, get an explicit temperature dependence.

\section{THERMOMAGNETIC RENORMALONS}\label{thermomagneticrenormalons}
In this section we are going to calculate the renormalon contribution taking finite temperature and also  the presence of a weak external magnetic field. We want to stress that for a strong magnetic field, we did not find renormalons. For our calculation we need the magnetic bosonic propagator in the weak field region $eB \ll m^{2}$, which is given by \cite{gluon}
\begin{eqnarray}\label{propmagneticodebil}
iD^B(k)&&\xrightarrow{eB\rightarrow 0} \frac{i}{k_\parallel^2 - k_\perp^2 -m^2} - \frac{i(eB)^2}{(k_\parallel^2 - k_\perp^2 -m^2)^3} \nonumber \\
&-& \frac{2i(eB)^2k_\perp^2}{(k_\parallel^2 - k_\perp^2 -m^2)^4}.
\end{eqnarray}
The corresponding expression for the renormalon type diagram  is now given by 
\begin{align}
R_{B,T,k}=\frac{1}{(-i\lambda)^{k-2}} \int \frac{d^4q}{(2\pi)^4} i D(p+q)[B^{(T,B)}(q)]^{k-1},
\end{align}
where $(T,B)$ refers to finite temperature and weak magnetic field effects. Following the same procedure,  let us first consider one bubble
\begin{eqnarray}
B^{(T,B)}(q)	&=&\frac{(-i\lambda)^2}{2} iT\sum_n\int\frac{d^3k}{(2\pi)^3}\; i\widetilde{D}^{(T,B)}(k)\; \nonumber \\
&\times& i\widetilde{D}^{(T,B)}(k+q),
\label{uno}
\end{eqnarray}
where $i\widetilde{D}^{(T,B)}(k)$ is is the finite temperature propagator, up to order $(eB)^2$ in the magnetic field, defined as
\begin{align}
i\widetilde{D}^{(T,B)}(k)  &\equiv iD^{T}(k)+iD^{(T,B)}(k),     
\end{align}
with 
\begin{eqnarray}
    iD^{T}(k)&=&-\frac{i}{\omega_n^2+k^2+m^2} \hspace{1cm}\text{and} \nonumber \\
iD^{(T,B)}(k)&=&\frac{i(eB)^2}{(\omega_n^2+k^2+m^2)^3} \nonumber \\
&-&\frac{2i(eB)^2\;k_{\perp}^2}{(\omega_n^2+k^2+m^2)^4}.
\end{eqnarray}
Therefore, using this notation Eq.(\ref{uno}) becomes
\begin{align}
\begin{split}
B^{(T,B)}(q)	=&\frac{(-i\lambda)^2}{2} iT\sum_n\int\frac{d^3k}{(2\pi)^3}\\ &\times(iD^{T}(k)+iD^{(T,B)}(k)) \\
&\times (iD^{T}(k+q)+iD^{(T,B)}(k+q)),  
\label{dos}
\end{split}
\end{align}
Note that the previous multiplication will produce terms of order greater than $(eB)^2$. If we restrict ourselves up to order $(eB)^2$, we obtain 
\begin{equation}
    B^{(T,B)}(q)	= D_1(k,q)+ D_2(k,q) + D_3(k,q),
\end{equation}
with 
\begin{align}
D_1(k,q)	&=\frac{(-i\lambda)^2}{2} iT \sum_n\int\frac{d^3k}{(2\pi)^3}iD^{T}(k)\;iD^{T}(k+q),\\
D_2(k,q)	&=\frac{(-i\lambda)^2}{2} iT \sum_n\int\frac{d^3k}{(2\pi)^3}iD^{T}(k)\;iD^{(T,B)}(k+q),\\
D_3(k,q)	&=\frac{(-i\lambda)^2}{2} iT \sum_n\int\frac{d^3k}{(2\pi)^3}iD^{(T,B)}(k)\;iD^{T}(k+q).
\end{align}
It is straightforward to prove that $D_2(k,q)=D_3(k,q)$.
Notice that $D_1(k,q)$ is the bubble with only temperature corrections obtained in the previous section (Eq.~(\ref{fishT})), hence 
\begin{align}
D_1(k,q)	&=\frac{-i\lambda^2}{32\pi^2}\left[\log\left(\frac{q^2}{\mu^2}\right)-\frac{8}{q^2}\sum_{n=1}^\infty \frac{m T}{n} K_1(nm/T)\right].
\end{align}
Now we have to calculate $D_3(k,q)$. Since we know that $D_2(k,q)=D_3(k,q)$, we have 
\begin{align}
D_3(k,q)=&\frac{(-i\lambda)^2}{2} iT \sum_n\int\frac{d^3k}{(2\pi)^3}iD^{(T,B)}(k)\;iD^{T}(k+q) \nonumber \\
=& \frac{(-i\lambda)^2}{2} iT \sum_n\int\frac{d^3k}{(2\pi)^3} \nonumber\\
&\times \Biggl[ \frac{i(eB)^2}{(\omega_n^2+k^2+m^2)^3} -\frac{2i(eB)^2\;k_{\perp}^2}{(\omega_n^2+k^2+m^2)^4}  \Biggr] \nonumber\\
&\times \Biggl[ -\frac{i}{(\omega_n+\omega)^2+(k+q)^2+m^2} \Biggr], \nonumber \\
\equiv &D_{3.1}(k,q) + D_{3.2}(k,q)
\end{align}
with
\begin{align}
\begin{split}
D_{3.1}(k,q)=&  \frac{(-i\lambda)^2 i T}{2} \sum_n\int\frac{d^3k}{(2\pi)^3}  \frac{(eB)^2}{(\omega_n^2+k^2+m^2)^3} \\&\times \Biggl[ \frac{1}{(\omega_n+\omega)^2+(k+q)^2+m^2} \Biggr], 
\end{split}
\end{align}
and
\begin{align}
\begin{split}
D_{3.2}(k,q)=&\frac{(-i\lambda)^2 i T}{2}\sum_n\int\frac{d^3k}{(2\pi)^3} \frac{2(eB)^2\;k_{\perp}^2}{(\omega_n^2+k^2+m^2)^4} \\
&\times \Biggl[ \frac{-1}{(\omega_n+\omega)^2+(k+q)^2+m^2} \Biggr] .
\end{split}
\end{align}
Let's first calculate $D_{3.1}(k,q)$, 
\begin{eqnarray}
 &&D_{3.1}(k,q)=  \frac{(-i\lambda)^2}{2\cdot 2!}\left(\frac{\partial}{\partial \tilde{m}^2}\right)^2 (eB)^2\;i T\sum_{n}\int\frac{d^3k}{(2\pi)^3} \nonumber \\
 &\times& \Biggl[ \frac{1}{\omega_n^2+k^2+\tilde{m}^2} \Biggr] 
 \Biggl[ \frac{1}{(\omega_n+\omega)^2+(k+q)^2+m^2} \Biggr], \label{d31}
\end{eqnarray}
where we have used
\begin{equation}
    \frac{1}{2!}\left(\frac{\partial}{\partial \tilde m^2}\right)^2\frac1{k^2+\tilde m^2}	=\frac1{(k^2+\tilde m^2)^3}. \label{derivadamasa1}
\end{equation}
We have again the expression found in the previous section (Eq.~(\ref{26})), with the difference that we have to derive with respect to $\tilde{m}^2$ twice. In this way, we obtain
\begin{eqnarray}
D_{3.1}(k,q)&=&  \frac{(-i\lambda)^2}{32 \pi^2} (eB)^2 \Biggl[ \frac{1}{2 m^2 q^2} \nonumber \\
&&-\frac{4}{q^2} \sum_{n=1}^\infty \frac{n}{4mT} K_1(nm/T).
\end{eqnarray}
In a similar fashion we calculate $D_{3.2}(k,q)$,
\begin{eqnarray}
 &&D_{3.2}(k,q)=  \frac{(-i\lambda)^2}{3!}\left(\frac{\partial}{\partial \tilde{m}^2}\right)^3 (eB)^2\;i T\sum_{n}\int\frac{d^3k}{(2\pi)^3} \nonumber \\
 &\times& \Biggl[ \frac{k_{\perp}^2}{\omega_n^2+k^2+\tilde{m}^2} \Biggr] 
 \Biggl[ \frac{-1}{(\omega_n+\omega)^2+(k+q)^2+m^2} \Biggr], \label{d32}
\end{eqnarray}
where we have  used
\begin{equation}
    -\frac{1}{3!}\left(\frac{\partial}{\partial \tilde m^2}\right)^3\frac1{k^2+\tilde m^2}=\frac1{(k^2+\tilde m^2)^4}. \label{derivadamasa2}
\end{equation}
Here we have once again the expression given in the previous section (Eq.~(\ref{26})), with the difference that we have to derive three times with respect to $\tilde{m}^2$. Hence, we obtain
\begin{eqnarray}
&&D_{3.2}(k,q)	=\frac{(eB)^2(-i\lambda)^2\; i}{2\cdot 3!}\nonumber
\\&&\times\left[\frac{-3!}{96\pi^2 m^2 q^2}+\frac{2}{6\pi^2 q^2} \sum_{n=1}^{\infty}\frac{3n}{8mT}K_1(nm/T) \right].
\end{eqnarray}Taking into account the results $D_1(k,q)$, $D_2(k,q)$ and $D_3(k,q)$, we obtain for $B^{(T,B)}(q)$ 
\begin{eqnarray}
&&B^{(T,B)}(q)	=\frac{-i\lambda^2}{32\pi^2} \Biggl[ \log \left(\frac{q^2}{\mu^2}\right)-\frac{8}{q^2}\sum_{n=1}^\infty \frac{m T}{n} K_1\left(\frac{nm}{T}\right) \nonumber \\ && 	+\frac{(eB)^2}{q^2} \Biggl(\frac{2}{3m^2} - \sum_{n=1}^\infty \frac{n}{mT} K_1\left(\frac{nm}{T}\right) 	\Biggr)\Biggr].
\end{eqnarray}

Following the same procedure of the previous section, we have to calculate the renormalon-type diagram
\begin{align}
R_{B,T,k}=\frac{1}{(-i\lambda)^{k-2}} \int \frac{d^4q}{(2\pi)^4} i D(p+q)[B^{(T,B)}(q)]^{k-1},
\end{align}
obtaining
\begin{eqnarray} 
R_{B,T,k} &=& -i \widetilde{m}^2 \left( \frac{\lambda}{32\pi ^2}  \right)^k \Gamma(k) +i \left( \frac{\lambda}{64\pi ^2} \right)^k \frac{m^2}{\mu^2}\widetilde{m}^2 \Gamma (k) \nonumber \\
 &-&2i \frac{\widetilde{m}^2}{\mu^2} \left( \frac{\lambda}{64\pi ^2}  \right)^k \Gamma(k) \Bigg[8\sum_{n=1}^\infty \frac{m T}{n} K_1\left(\frac{nm}{T}\right) \nonumber \\ && 	+(eB)^2 \Biggl(\frac{2}{3m^2} - \sum_{n=1}^\infty \frac{n}{mT} K_1\left(\frac{nm}{T}\right) 	\Biggr) \Biggr] 
\end{eqnarray}
This can be written as
\begin{eqnarray}
R_{B,T,k}&\equiv& -i \widetilde{m}^2 \left( \frac{\lambda}{32\pi ^2}  \right)^k \Gamma(k) +i \left( \frac{\lambda}{64\pi ^2} \right)^k \frac{m^2}{\mu^2}\widetilde{m}^2 \Gamma (k) \nonumber \\
 &-&2i \frac{\widetilde{m}^2}{\mu^2} \left( \frac{\lambda}{64\pi ^2}  \right)^k \Gamma(k) F_{\text{B,T}} + ... \hspace{1mm}  \text{,}
\end{eqnarray}
where
\begin{eqnarray}
F_{\text{B,T}}&=& -8\sum_{n=1}^\infty \frac{m T}{n} K_1\left(\frac{nm}{T}\right) \nonumber \\ && 	+(eB)^2 \Biggl(\frac{2}{3m^2} - \sum_{n=1}^\infty \frac{n}{mT} K_1\left(\frac{nm}{T}\right) 	\Biggr).   \label{fbt} \end{eqnarray}
Now we can find the Borel transform $B[b]$
\begin{equation}
B[b] = \frac{-i \widetilde{m}^2}{b-32\pi ^2} +\Biggl[i \frac{m^2}{\mu^2}\widetilde{m}^2- \frac{2i \widetilde{m}^2}{ \mu^2}F_{\text{B,T}} \Biggr] \frac{1}{b-64\pi ^2} + ...  \text{.} 
\end{equation}
As in the pure thermal situation, the location of the poles does not depend in this case on temperature and on the strength of the magnetic field. The residue, however,becomes thermomagnetic dependent.

\section{THERMOELECTRIC RENORMALONS}\label{thermoelectricrenormalons}
Thermoelectric corrections will be handled using an electric field dependent finite temperature bosonic propagator. We are going to introduce the propagator taking first only the external electric field into account, incorporating then, later, finite temperature effects. The bosonic propagator in Euclidean space is given by \cite{nosotros}
\begin{equation}D(k) =  \int _{0} ^{\infty}{ds}\frac{e^ {-s\left(\frac{\tanh (eEs)}{eEs} k_{\parallel}^2 +k_{\perp}^2+ m^2 \right)}}{\cosh (eEs)}, \label{propE}
\end{equation}
where $e$ is the electric charge, $k_{\parallel}$ and $k_{\perp}$
refer to $(k_{4},0, 0, k_{3})$ and $(0, k_{1},k_{2},0)$, respectively. Note that in the euclidean version $k^{2} = k_{\parallel}^2 + k_{\perp}^2$=$k_4^2+k_3^2+k_1^2+k_2^2$. Since we will be interested in the weak electric field region where $eE \ll m^{2}$, by expanding the previous expression we get up to order  $\mathcal{O}(E^2)$ \cite{nosotros}
\begin{eqnarray}
iD^{E}(k)&\approx& \frac{i}{k^2+m^2} +i(eE)^2\Biggl(-\frac{1}{(k^2+m^2)^3} \nonumber \\
&&+\frac{2  k_{\parallel}^2}{(k^2+m^2)^4}\Biggr) \nonumber \\
&\approx& \frac{i}{k^2+m^2} +i(eE)^2\Biggl(\frac{1}{(k^2+m^2)^3} \nonumber \\
&&-\frac{2(k_{\perp}^2+m^2)  }{(k^2+m^2)^4}\Biggr)  . \label{debil7}
\end{eqnarray}
The corresponding expression for the renormalon type diagram  is now given by 
\begin{align}
R_{E,T}=\frac{1}{(-i\lambda)^{k-2}} \int \frac{d^4q}{(2\pi)^4} i D(p+q)[B^{(T,E)}(q)]^{k-1},
\end{align}
where $(E,B)$ refers to finite temperature and weak electric field effects. Following the same procedure presented in the previous sections let us first consider one bubble
\begin{eqnarray}
B^{(T,E)}(q)	&=&\frac{(-i\lambda)^2}{2} iT\sum_n\int\frac{d^3k}{(2\pi)^3}\; i\widetilde{D}^{(T,E)}(k)\; \nonumber \\
&\times& i\widetilde{D}^{(T,E)}(k+q),
\label{uno2}
\end{eqnarray}
where $i\widetilde{D}^{(T,B)}(k)$ is is the finite temperature propagator, up to order $(eB)^2$ in the magnetic field, given by 
\begin{align}
i\widetilde{D}^{(T,E)}(k)  &\equiv iD^{T}(k)+iD^{(T,E)}(k),    
\end{align}
with 
\begin{eqnarray}
    iD^{T}(k)&=&-\frac{i}{\omega_n^2+k^2+m^2} \hspace{1cm}\text{and} \nonumber \\
iD^{(T,E)}(k)&=&\frac{i(eE)^2}{(\omega_n^2+k^2+m^2)^3} \nonumber \\
&-&\frac{2i(eE)^2(k_{\perp}^2+m^2)}{(\omega_n^2+k^2+m^2)^4}.
\end{eqnarray}
Therefore, using this notation Eq.(\ref{uno2}) becomes
\begin{align}
\begin{split}
B^{(T,E)}(q)	=&\frac{(-i\lambda)^2}{2} iT\sum_n\int\frac{d^3k}{(2\pi)^3}\\ &\times(iD^{T}(k)+iD^{(T,E)}(k)) \\
&\times (iD^{T}(k+q)+iD^{(T,E)}(k+q)),  
\label{dos2}
\end{split}
\end{align}
Note that the previous multiplication will produce terms of order greater than $(eE)^2$. If we restrict ourselves up to order $(eE)^2$, we obtain 
\begin{equation}
    B^{(T,E)}(q)	= D_1(k,q)+ D_2(k,q) + D_3(k,q),
\end{equation}
with 
\begin{align}
D_1(k,q)	&=\frac{(-i\lambda)^2}{2} iT \sum_n\int\frac{d^3k}{(2\pi)^3}iD^{T}(k)\;iD^{T}(k+q),\\
D_2(k,q)	&=\frac{(-i\lambda)^2}{2} iT \sum_n\int\frac{d^3k}{(2\pi)^3}iD^{T}(k)\;iD^{(T,E)}(k+q),\\
D_3(k,q)	&=\frac{(-i\lambda)^2}{2} iT \sum_n\int\frac{d^3k}{(2\pi)^3}iD^{(T,E)}(k)\;iD^{T}(k+q).
\end{align}
It is straightforward to prove that $D_2(k,q)=D_3(k,q)$.
Notice that $D_1(k,q)$ is the bubble with only temperature corrections obtained in the previous section (Eq.~(\ref{fishT})), hence 
\begin{align}
D_1(k,q)	&=\frac{-i\lambda^2}{32\pi^2}\left[\log\left(\frac{q^2}{\mu^2}\right)-\frac{8}{q^2}\sum_{n=1}^\infty \frac{m T}{n} K_1(nm/T)\right].
\end{align}
Now we have to calculate $D_3(k,q)$. Since we know that $D_2(k,q)=D_3(k,q)$, we have 
\begin{align}
D_3(k,q)=&\frac{(-i\lambda)^2}{2} iT \sum_n\int\frac{d^3k}{(2\pi)^3}iD^{(T,E)}(k)\;iD^{T}(k+q) \nonumber \\
=& \frac{(-i\lambda)^2}{2} iT \sum_n\int\frac{d^3k}{(2\pi)^3} \nonumber\\
&\times \Biggl[ \frac{i(eE)^2}{(\omega_n^2+k^2+m^2)^3} -\frac{2i(eE)^2(k_{\perp}^2+m^2)}{(\omega_n^2+k^2+m^2)^4}  \Biggr] \nonumber\\
&\times \Biggl[ -\frac{i}{(\omega_n+\omega)^2+(k+q)^2+m^2} \Biggr], \nonumber \\
\equiv &D_{3.1}(k,q) + D_{3.2}(k,q)
\end{align}
with
\begin{align}
\begin{split}
D_{3.1}(k,q)=&  \frac{(-i\lambda)^2 i T}{2} \sum_n\int\frac{d^3k}{(2\pi)^3}  \frac{(eE)^2}{(\omega_n^2+k^2+m^2)^3} \\&\times \Biggl[ \frac{1}{(\omega_n+\omega)^2+(k+q)^2+m^2} \Biggr], 
\end{split}
\end{align}
and
\begin{align}
\begin{split}
D_{3.2}(k,q)=&\frac{(-i\lambda)^2 i T}{2}\sum_n\int\frac{d^3k}{(2\pi)^3} \frac{2(eE)^2(k_{\perp}^2+m^2)}{(\omega_n^2+k^2+m^2)^4} \\
&\times \Biggl[ \frac{-1}{(\omega_n+\omega)^2+(k+q)^2+m^2} \Biggr]. 
\end{split}
\end{align}
Let's first calculate $D_{3.1}(k,q)$, 
\begin{eqnarray}
 &&D_{3.1}(k,q)=  \frac{(-i\lambda)^2}{2\cdot 2!}\left(\frac{\partial}{\partial \tilde{m}^2}\right)^2 (eE)^2\;i T\sum_{n}\int\frac{d^3k}{(2\pi)^3} \nonumber \\
 &\times& \Biggl[ \frac{1}{\omega_n^2+k^2+\tilde{m}^2} \Biggr] 
 \Biggl[ \frac{1}{(\omega_n+\omega)^2+(k+q)^2+m^2} \Biggr], 
\end{eqnarray}
where we have used the relation of Eq.~(\ref{derivadamasa1}). This expression is analogue of Eq.~(\ref{d31}). Hence, we obtain
\begin{eqnarray}
D_{3.1}(k,q)&=&  \frac{(-i\lambda)^2}{32 \pi^2} (eE)^2 \Biggl[ \frac{1}{2 m^2 q^2} \nonumber \\
&&-\frac{4}{q^2} \sum_{n=1}^\infty \frac{n}{4mT} K_1(nm/T).
\end{eqnarray}
In a similar way we calculate $D_{3.2}(k,q)$,
\begin{eqnarray}
 &&D_{3.2}(k,q)=  \frac{(-i\lambda)^2}{3!}\left(\frac{\partial}{\partial \tilde{m}^2}\right)^3 (eE)^2\;i T\sum_{n}\int\frac{d^3k}{(2\pi)^3} \nonumber \\
 &\times& \Biggl[ \frac{(k_{\perp}^2+m^2)}{\omega_n^2+k^2+\tilde{m}^2} \Biggr]  \Biggl[ \frac{-1}{(\omega_n+\omega)^2+(k+q)^2+m^2} \Biggr], 
\end{eqnarray}
where we have used the relation of Eq.~(\ref{derivadamasa2}). This expression is analogue of Eq.~(\ref{d32}). Therefore, we obtain
\begin{eqnarray}
 &&D_{3.2}(k,q)=  \frac{(-i\lambda)^2}{32 \pi^2} (eE)^2 \Biggl[ -\frac{1}{m^2 q^2} \nonumber \\
 &&-\frac{1}{3q^2} \sum_{n=1}^\infty \frac{n}{mT} K_1(nm/T) +\frac{1}{3}\sum_{n=1}^\infty \frac{n^2}{T^2} K_2(nm/T)\Biggr]. \nonumber \\
\end{eqnarray}
Following an analogous procedure to the previous sections, we obtain for the renormalon-type diagram 
\begin{eqnarray}
R_{E,T,k}&\equiv& -i \widetilde{m}^2 \left( \frac{\lambda}{32\pi ^2}  \right)^k \Gamma(k) +i \left( \frac{\lambda}{64\pi ^2} \right)^k \frac{m^2}{\mu^2}\widetilde{m}^2 \Gamma (k) \nonumber \\
 &-&2i \frac{\widetilde{m}^2}{\mu^2} \left( \frac{\lambda}{64\pi ^2}  \right)^k \Gamma(k) F_{\text{E,T}} + ... \hspace{1mm}  \text{,}
\end{eqnarray}
where
\begin{eqnarray}
F_{\text{E,T}}&=& -8\sum_{n=1}^\infty \frac{m T}{n} K_1\left(\frac{nm}{T}\right) \nonumber \\ && 	+(eE)^2 \Biggl(-\frac{2}{m^2} - \sum_{n=1}^\infty \frac{2n}{3mT} K_1\left(\frac{nm}{T}\right) \nonumber \\
&&+ \sum_{n=1}^\infty \frac{2n^2}{3T^2} K_2\left(\frac{nm}{T}\right) 	\Biggr).   \label{fet} \end{eqnarray}
Now we can find the Borel transform $B[b]$
\begin{equation}
B[b] = \frac{-i \widetilde{m}^2}{b-32\pi ^2} +\Biggl[i \frac{m^2}{\mu^2}\widetilde{m}^2- \frac{2i \widetilde{m}^2}{ \mu^2}F_{\text{E,T}} \Biggr] \frac{1}{b-64\pi ^2} + ...  \text{.} 
\end{equation}
\noindent  We notice that again the position of the poles does not depend on temperature and on the electric field as well. However, the residue does get a thermoelectric dependence. Further, as it occurred in the thermomagnetic case, the  strong electric field limit does not have any effects on the residue of the  renormalons.

\section{RESULTS}\label{conclusions}
\begin{figure}[h]
\includegraphics[width=88mm]{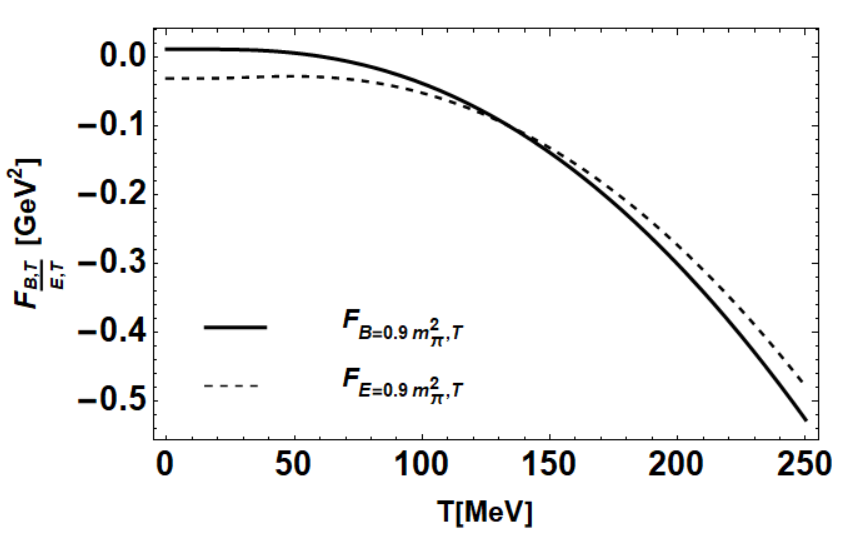} 
\caption{Thermal evolution of the thermomagnetic  ($F_{\text{E,T}}$) and thermoelectric  ($F_{\text{E,T}}$) residues for given values  $eB=0.9 m_\pi^2$ and $eE=0.9 m_\pi^2$ of the magnetic and electric field intensities.}
\label{residuo1}
\end{figure}

\begin{center}
\begin{figure}[h]
\includegraphics[width=18cm]{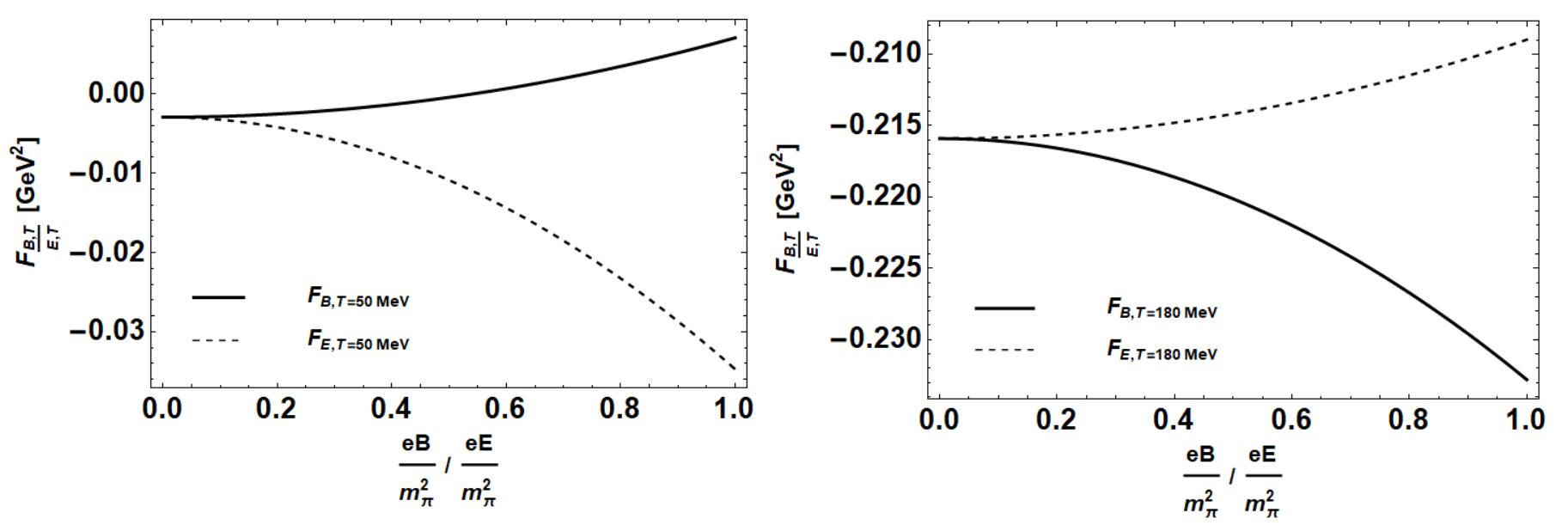} 
\caption{Thermomagnetic and thermoelectric residues ($F_{\text{E,T}}$) and ($F_{\text{E,T}}$) as function of the field intensities. The left panel is  for $T = 50$ MeV and the right panel for $T = 180$ MeV. }
\label{residuo2}
\end{figure}
\end{center}

According to our previous discussion, the interesting quantity to be analyzed is the evolution of the thermomagnetic and thermoelectric  residues  ($F_{\text{B,T}}$), defined in Eq.~(\ref{fbt}),  and ($F_{\text{E,T}}$) defined en Eq.~(\ref{fet}), respectively. 
In Fig.~\ref{residuo1} we show the  thermomagnetic and thermoelectric evolution of renormalon residues  ($F_{\text{E,T}}$) and ($F_{\text{E,T}}$) as function of temperature for  given strengths $eB=0.9 m_\pi^2$  and $eE=0.9 m_\pi^2$ of the magnetic and electric fields, respectively. We notice that these residues have a similar behavior, being both negative, although initially the thermoelectric renormalon has a more negative value than the magnetic case. For $T\sim130$ MeV this behavior changes being then the magnetic residue more negative than the electric case. In Fig.~\ref{residuo2} the thermomagnetic and thermoelectric residues are shown as function of their respective field intensities. The left right panel corresponds to $T = 50$ Mev whereas the right panel corresponds to $T = 180$ MeV. We notice that their behaviors are opposite. However, for  $T=50$ MeV we see that the thermomagnetic residue grows. When temperature starts to increase, this behavior changes until is starts to diminish. The same happens with the thermoelectric residue. It changes from an initially decreasing into a growing function. However, both residues have  opposite behaviors as function of their corresponding field strengths.
%
\onecolumngrid
\twocolumngrid

\section{CONCLUSIONS}\label{conclusions}

In this work we have analyzed the thermal behavior of renormalons in the $\lambda \phi^4$ theory including the presence of an external weak electric and a weak magnetic field,  taken both cases separately. Our analysis is valid for the whole range of temperature.
First we considered only temperature effects obtaining a closed analytic expression which coincides,  in the low and high temperature regions, with previous results reported in   \cite{renormalontermomagnetico} and \cite{Loewe10}, respectively. In a second step we went through the calculation  of thermomagnetic effects. The difference with previous results  \cite{renormalontermomagnetico} is again the expression for the temperature  dependence. Finally, we addressed the new issue of thermoelectric effects on renormalons. In this case we found a competition between the electric and magnetic effects, having opposite behaviors. Although there is an approximate temperature  $T\sim130$ MeV where the magnetic and electric field dependent residues, change  their behavior, they continue having this opposite dependence on the corresponding field strengths. In a future work we will discuss the simultaneous effect of both fields, i.e. a thermomagnetic-electric scenario. As a final general comment about our calculation of thermo-magnetic and thermo-electric effects associated to the renormalon diagram, it is important to stress that the leading renormalon behavior is associated to the vacuum contribution. Thermal effects, for example, affect only the residue of sub-leading poles. In the case of the influence of external magnetic or electric fields this conclusion remains valid as well. If we would have attached the external effects in the single propagator of our diagram, instead of considering  these effects only attached to  the chain of bubbles, no interference effects will appears between both diagrams that could affect the leading term. This point has been mentioned in \cite{malaquias}. Finally, thinking about  possible observable consequences, we would like to consider in a future work renormalons effects in the determination of $\pi$-$\pi$ scattering lengths described from the perspective of our model.

\section*{Acknowledgements}
M. Loewe and R. Zamora acknowledge support from   ANID/CONICYT FONDECYT Regular (Chile) under Grant No. 1200483. M.L.  acknowledges support from Fondecyt under grants No. 1190192 and No. 1170107. ML acknowledges also support from Anid/Pia/Basal (Chile) under grant No. FB082




\end{document}